\documentclass{article}
\usepackage{amsfonts}

\newcommand{\myfrac}[2]{\frac{\textstyle #1 }{\textstyle #2 }}
\newcommand{\ep}[1]{ \, \mathbb{E}_{#1} }
\newcommand{\en}[1]{ \, \overline{\mathbb{E}}_{#1} }
\newcommand{\unitmatrix}{ \,\mathbb{I} }

\makeatletter \@addtoreset{equation}{section} \makeatother
\renewcommand{\theequation}{\thesection.\arabic{equation}}

\begin{document}

\title{Functional representation of the Volterra hierarchy.}

\author{V.E. Vekslerchik
  \\[10mm]
Universidad de Castilla-La Mancha, Ciudad Real, Spain
  \\
and
  \\
Institute for Radiophysics and Electronics, Kharkov, Ukraine
  \\[10mm]
\texttt{E-mail: vadym@ind-cr.uclm.es}
}

\date{\today}

\maketitle

\begin{abstract}
\noindent 
In this paper I study the functional representation of the 
Volterra hierarchy (VH). Using the Miwa's shifts I rewrite the 
infinite set of Volterra equations as one functional equation. 
This result is used to derive a formal solution of the 
associated linear problem, a generating function for the 
conservation laws and to obtain a new form of the Miura and 
Backlund transformations. I also discuss some relations between 
the VH and KP hierarchy.
\end{abstract}

\section{Introduction.}  \label{sec-intro}

In this paper I want to discuss an application of the so-called 
functional equation approach to one of the oldest integrable 
discrete systems, namely the Volterra model,
\begin{equation}
  \dot{u}_{n} = u_{n} \left( u_{n+1} - u_{n-1} \right).
\label{ve-dot}
\end{equation}
where the dot stands for the differentiating with respect to time.
It was proposed many years ago for the description of the 
population dynamics \cite{L1910,L1920,V}. Later it was applied to 
many physical phenomena, such as, e.g., collapse of Langmuir 
waves, nonlinear LC nets, Liouville field theory. It is known to 
be integrable since the paper by Manakov \cite{M} (see also 
\cite{KM}) who developed the corresponding version of the 
inverse scattering transform (IST).

The IST is a method which drastically changed the theory of PDEs, 
of nonlinear systems as well as many other fields of nonlinear 
mathematics and physics. However the practical implementation of 
its algorithms is not so easy as one might expect. The results 
which can be obtained using the IST are usually formulated in 
terms of some auxiliary objects (Jost functions, which are 
solutions of some auxiliary linear problems, and scattering data, 
which link different Jost functions) and the main difficulty of 
this approach is that one cannot find them explicitly or to 
express them in terms of the $u_{n}$. As a result, for example, 
one cannot get closed expression for the generating function for 
the constants of motion, but only an algorithm how to derive it.
That is why during all the years of the modern theory of 
integrable systems people were looking for some other tools to 
deal with these particular equations which are called integrable. 
One of such approaches is the topic of this work. In a few words 
it can be described as follows: instead of your equation 
(Volterra equation in our case) you consider an infinite family 
of similar equations (Volterra hierarchy in our case) and then 
instead of an infinite number of differential equations you deal 
with one (or a few) equation of other kind, functional or 
difference one. 

Starting from the standard IST representation of the VH, I derive 
in section \ref{sec-zcr} the functional equations for the 
tau-functions of the VH. After discussing the superposition 
formulae for different Miwa's shifts (section \ref{sec-fay}) I 
obtain in section \ref{sec-baf} a formal solution of the 
auxiliary linear problem. Using these results I derive the 
generating function for the conservation laws of the VH (section 
\ref{sec-cl}), address the problem of Miura and Backlund 
transformations (section \ref{sec-bt}) and obtain in section 
\ref{sec-sls} dark-soliton solutions of the Volterra equations. 
Finally, I present some results on interrelations between the 
Volterra and some other hierarchies (section \ref{sec-kp}). 

\section{Zero-curvature representation and Miwa's shifts.} \label{sec-zcr}

The inverse scattering approach for integrable systems is based on
the zero-curvature representation (ZCR), when our nonlinear 
equations are presented as a compatibility condition for some 
linear system. For the VH this system can be written as \cite{FT}
\begin{equation}
  \left\{
  \begin{array}{lcl}
  \Psi_{n+1} & = & U_{n} \Psi_{n}
  \cr
  \dot\Psi_{n} & = & V_{n} \Psi_{n}
  \end{array}
  \right.
\label{zcr-system}
\end{equation}
where $\Psi_{n}$ is a $2$-column (or $2 \times 2$ matrix),
\begin{equation}
  U_{n} = \pmatrix{ \lambda & u_{n} \cr -1 & 0 }
\label{zcr-u}
\end{equation}
and $V_{n}$ is some $2 \times 2$ matrix depending on the quantities 
$u_{n}$, $u_{n \pm 1}$, ... as well as on the auxiliary parameter $\lambda$

Traditionally Volterra chains are considered more often in the 
framework of the 'big' Lax representation, when the system is 
finite ($n=1,...,N$) and $U$ and $V$ are $N \times N$ matrices. In 
this paper we will use the $2 \times 2$ $U$-$V$ pair. This 
approach is equivalent to the $N \times N$ representation and 
sometimes even more convenient (say, it can be more easily 
modified to the cases of different boundary conditions, including 
the soliton case when $N=\infty$).

To provide the self-consistency of the system (\ref{zcr-system}) 
the matrices $V_{n}$ have to satisfy the 
following equation
\begin{equation}
  \dot U_{n} = V_{n+1} U_{n} - U_{n} V_{n}. 
\label{zcr-eq}
\end{equation}

The choice of the $V$-matrix for the given $U$-matrix, is not unique.
One can find an infinite number of matrices $V$, which are polynomials 
of different order in $\lambda$, such that ZCR
(\ref{zcr-eq}) will be satisfied for all values of $\lambda$ provided
the functions $u_{n}$ solve some nonlinear evolutionary equations. 
These equations are called 'higher Volterra equations'. Taken together
they constitute the VH.

Using the notation
\begin{equation}
  V_{n} = \pmatrix{ a_{n} & b_{n} \cr c_{n} & d_{n} }
\end{equation}
one can rewrite (\ref{zcr-eq}) as
\begin{eqnarray}
  0 &=& \lambda \left( a_{n+1} - a _{n} \right) - b_{n+1} - u_{n}c_{n}
  \\
  0 &=& b_{n} + u_{n}c_{n+1}
  \\
  0 &=& \lambda c_{n+1} + a _{n} - d_{n+1}
\end{eqnarray}
and
\begin{equation}
  \dot u_{n} = u_{n} \left( a_{n+1} - d _{n} \right) - \lambda b_{n}
\end{equation}
or, after eliminating $b_{n}$ and $d_{n}$,
\begin{equation}
  0 =
    \lambda \left( a_{n+1} - a _{n} \right) +
    u_{n+1} c_{n+2} - u_{n}c_{n},
\label{zcr-1a}
\end{equation}
and
\begin{equation}
  \dot u_{n} =
  u_{n} \left[ 
    a_{n+1} - a_{n-1} + \lambda \left( c_{n+1} - c _{n} \right)
  \right].
\label{zcr-1b}
\end{equation}
It is easy to show that (\ref{zcr-1a}) and (\ref{zcr-1b}) possess 
solutions where $a_{n}$'s are polynomials of the $(2j-2)$th order 
while $c_{n}$'s are polynomials of the $(2j)$th order in 
$\lambda$ for $j=1,2,...$. In what follows I indicate different 
polynomials with the upper index, $a_{n}^{(j)}, c_{n}^{(j)}$, and 
introduce an infinite set of times, $t_{j}$, to distinguish 
the resulting nonlinear equations.

By simple algebra one can establish the following relations
between different polynomials:
\begin{eqnarray}
  a_{n}^{(j+1)} & = & \lambda^{2} a_{n}^{(j)} + \alpha_{n}^{(j)},
\label{def-alpha}
\\
  c_{n}^{(j+1)} & = & \lambda^{2} c_{n}^{(j)} + \lambda\gamma_{n}^{(j)}
\label{def-gamma}
\end{eqnarray}
where $\alpha_{n}^{(j)}$ and $\gamma_{n}^{(j)}$ do not depend on
$\lambda$. Substituting (\ref{def-alpha}) and (\ref{def-gamma}) 
into (\ref{zcr-1a}) and (\ref{zcr-1b}) one can convert our system 
into
\begin{eqnarray}
&&
  \alpha_{n}^{(j)} + \alpha_{n-1}^{(j)} + \gamma_{n}^{(j+1)} 
  = 0,
\label{zcr-gamma}
\\
&&
  \alpha_{n+1}^{(j)} - \alpha_{n}^{(j)} 
  + u_{n+1} \gamma_{n+2}^{(j)} - u_{n} \gamma_{n}^{(j)}
  = 0
\label{zcr-alpha}
\end{eqnarray}
and
\begin{equation}
  \partial_{j} u_{n} = 
  u_{n} \left[ \gamma_{n}^{(j)} - \gamma_{n+1}^{(j)} \right]
  =
  u_{n} \left[ \alpha_{n+1}^{(j-1)} - \alpha_{n-1}^{(j-1)} \right]
\label{zcr-dju}
\end{equation}
with $\partial_{j} = \partial / \partial t_{j}$

The simplest solution of (\ref{zcr-gamma}) and (\ref{zcr-alpha}),
\begin{equation}
  \alpha^{(0)}_{n} = u_{n},
\qquad
\qquad
  \gamma^{(0)}_{n} = -1
\end{equation}
leads to the classical Volterra equation:
\begin{equation}
  \partial_{1} u_{n} =  
  u_{n} \left( u_{n+1} - u_{n-1} \right)
\label{ve-1}
\end{equation}
which can be rewritten as
\begin{equation}
  \tau_{n-1} \partial_{1} \tau_{n} - 
  \tau_{n}   \partial_{1} \tau_{n-1} = 
  \tau_{n+1} \tau_{n-2}
\label{blve-1}
\end{equation}
where the tau-functions of the VH, $\tau_{n}$, are defined by
\begin{equation}
  u_{n} = { \tau_{n+1} \tau_{n-2} \over \tau_{n} \tau_{n-1} }.
\label{def-tau}
\end{equation}

In terms of the tau-functions the quantities $\alpha^{(j)}_{n}$ 
and $\gamma^{(j)}_{n}$ can be presented as
\begin{eqnarray}
  \alpha_{n}^{(j)} & = & 
  \partial_{j+1} \ln \frac{\tau_{n}}{\tau_{n-1}}
\label{alpha-as-dj}
\\
  \gamma_{n}^{(j)} & = & 
  \partial_{j} \ln \frac{\tau_{n-2}}{\tau_{n}}
\label{gamma-as-dj}
\end{eqnarray}
which makes (\ref{zcr-dju}) and (\ref{zcr-gamma}) to be satisfied 
automatically while equation (\ref{zcr-alpha}) (together with 
(\ref{blve-1})) becomes a recurrent formula for the Volterra 
flows:
\begin{equation}
  \partial_{j+1} \ln \frac{\tau_{n}}{\tau_{n-1}} =
  u_{n} \partial_{j} \ln \frac{\tau_{n}}{\tau_{n-2}} +
  \partial_{j}\partial_{1} \ln \tau_{n}. 
\label{vh-rec}
\end{equation}

Till now we were following the standard zero-curvature scheme, 
but hereafter, since our purpose is to discuss the VH as a whole, 
I will deal not with the quantities $\alpha^{(j)}_{n}$ and 
$\gamma^{(j)}_{n}$ (which describe the $j$th flow) but with 
series defined by
\begin{equation}
  \alpha_{n}(\zeta) = \sum_{j=0}^{\infty} \alpha^{(j)}_{n} \zeta^{j},
  \qquad
  \qquad
  \gamma_{n}(\zeta) = \sum_{j=0}^{\infty} \gamma^{(j)}_{n} \zeta^{j}
\end{equation}
and will consider the quantities $u_{n}$ (and $\tau_{n}$) as 
functions of an infinite set of times $\left\{ t_{j} 
\right\}_{j=1,2,...}$:
\begin{equation}
  u_{n} = 
  u_{n}\left( \mathrm{t} \right) =
  u_{n}\left( t_{1}, t_{2}, t_{3},... \right) 
\end{equation}
which is possible due to the fact that all flows of the Volterra 
hierarchy $\partial / \partial t_{j}$ commute.

Using $\alpha_{n}(\zeta)$ and $\gamma_{n}(\zeta)$ one can present 
the elements of the $V$-matrices describing different Volterra 
flows in a typical for the theory of integrable systems 
'multiply+project' way:
\begin{equation}
  a^{(j)}_{n}(\lambda) = 
    \biggl[ 
    \lambda^{2j-2} \alpha_{n}\left(\lambda^{-2}\right) 
    \biggr]_{\ge 0},
  \qquad
  \qquad
  c^{(j)}_{n}(\lambda) = 
    \biggl[ 
    \lambda^{2j-1} \gamma_{n}\left(\lambda^{-2}\right) 
    \biggr]_{\ge 0}
\end{equation}
with projection $[ ... ]_{\ge 0}$ being defined by
\begin{equation}
  \left[ 
    \sum_{j=-\infty}^{\infty} f_{j}\lambda^{j}
  \right]_{\ge 0} =
  \sum_{j=0}^{\infty} f_{j}\lambda^{j}.
\end{equation}

In terms of $\alpha_{n}(\zeta)$ and $\gamma_{n}(\zeta)$ 
and the differential operator
\begin{equation}
  \partial(\zeta) = \sum_{j=1}^{\infty} \zeta^{j} \partial_{j}
\end{equation}
equations (\ref{alpha-as-dj}), (\ref{gamma-as-dj}) become
\begin{eqnarray}
  \alpha_{n}(\zeta) & = & 
  \frac{1}{\zeta} \partial(\zeta) \ln\frac{\tau_{n}}{\tau_{n-1}}
\label{alpha-as-dd}
\\
  \gamma_{n}(\zeta) & = & 
  -1 + \partial(\zeta) \ln\frac{\tau_{n-2}}{\tau_{n}}
\end{eqnarray}
while equations (\ref{zcr-gamma}), (\ref{zcr-alpha}) and 
(\ref{vh-rec}) lead to
\begin{eqnarray}
&&
  1 + 
  \zeta\alpha_{n}(\zeta) + 
  \zeta\alpha_{n-1}(\zeta) +
  \gamma_{n}(\zeta) 
  = 0
\label{zcr-gamma-zeta}
\\&&
  \alpha_{n+1}(\zeta) - \alpha_{n}(\zeta) +
  u_{n+1} \gamma_{n+2}(\zeta) - u_{n} \gamma_{n}(\zeta)
  = 0
\label{zcr-alpha-zeta}
\end{eqnarray}
and
\begin{equation}
  \alpha_{n}(\zeta) =
  - u_{n} \gamma_{n}(\zeta) 
  + \partial_{1}\partial(\zeta) \ln\tau_{n}.
\label{dd-tau}
\end{equation}
By multiplying (\ref{zcr-alpha-zeta}) by $\gamma_{n+1}(\zeta)$ 
and using 
\begin{equation}
  \partial(\zeta) u_{n} =
  u_{n} \left[ \gamma_{n}(\zeta) - \gamma_{n+1}(\zeta) \right]
\end{equation}
one can derive an invariant of the map 
(\ref{zcr-gamma-zeta}) and (\ref{zcr-alpha-zeta}):
\begin{equation}
  \alpha_{n}(\zeta) + \zeta\alpha_{n}^{2}(\zeta) -
  u_{n} \gamma_{n}(\zeta) \gamma_{n+1}(\zeta) 
  = \mbox{constant}.
\end{equation}
Noting that $\alpha_{n}^{(j)}$ and $\gamma_{n}^{(j)}$ are homogeneous 
polynomials in $u_{n}$ of $(j+1)$th and $j$th order 
correspondingly one can conclude that the constant in the 
right-hand side of the last formula is zero,
\begin{equation}
  \alpha_{n}(\zeta) + \zeta\alpha_{n}^{2}(\zeta) -
  u_{n} \gamma_{n}(\zeta) \gamma_{n+1}(\zeta) 
  = 0.
\label{Delta}
\end{equation}

To proceed further we have to derive some differential identities 
which are satisfied by $\alpha_{n}(\zeta)$ and 
$\gamma_{n}(\zeta)$ as functions of the infinite set of times 
$t_{j}$. It is easy to note that the simplest of the Volterra 
equations, (\ref{ve-1}), gives for the derivative 
  $\partial_{1}\alpha_{n}^{(0)}$, 
  $\alpha_{n}^{(0)}=u_{n}$, 
the following expression:
\begin{equation}
  \partial_{1} \alpha_{n}^{(0)} = 
  u_{n} \left[ 
    \gamma_{n}^{(0)} \gamma_{n+1}^{(1)} -
    \gamma_{n}^{(1)} \gamma_{n}^{(0)} 
  \right].
\end{equation}
By simple algebra one can derive the similar expression for the 
first derivative of 
  $\alpha_{n}^{(1)} = u_{n}\left(u_{n-1}+u_{n}+u_{n-1}\right)$, 
\begin{equation}
  \partial_{1} \alpha_{n}^{(1)} = 
  u_{n} \left[ 
    \gamma_{n}^{(0)} \gamma_{n+1}^{(2)} -
    \gamma_{n}^{(2)} \gamma_{n}^{(0)} 
  \right].
\end{equation}
It turns out that these formulae are the simplest cases of a more 
general identity which in terms of the series $\alpha_{n}(\zeta)$ 
and $\gamma_{n}(\zeta)$ reads
\begin{equation}
  \partial(\eta) \alpha_{n}(\xi) = 
    { \eta \over \xi - \eta } \, u_{n} \,
  \left[
    \gamma_{n+1}(\xi) \gamma_{n}(\eta) -
    \gamma_{n}(\xi)   \gamma_{n+1}(\eta)
  \right].
\label{prop-1}
\end{equation}
A proof of (\ref{prop-1}) can be given as follows. First let us 
note that
\begin{equation}
  \eta \, \partial(\xi) \alpha_{n}(\eta) = 
  \xi  \, \partial(\eta)\alpha_{n}(\xi)
\label{d-alpha-symm}
\end{equation}
(as follows from (\ref{alpha-as-dd})). Then, application of 
$\partial(\xi)$ to (\ref{dd-tau}) leads to
\begin{equation}
  \partial(\xi)\alpha_{n}(\eta) = 
  u_{n} \gamma_{n+1}(\xi) \gamma_{n}(\eta) +
  S_{n}(\xi,\eta)
\end{equation}
where $S_{n}(\xi,\eta)$ is a symmetric function of $\xi$ and 
$\eta$,  $S_{n}(\xi,\eta)=S_{n}(\eta,\xi)$. Subtracting from this 
equation a similar one with $\xi$ and $\eta$ being interchanged 
one can get
\begin{equation}
  \partial(\xi) \alpha_{n}(\eta) -
  \partial(\eta)\alpha_{n}(\xi) =
  u_{n} \,
  \left[
    \gamma_{n+1}(\xi) \gamma_{n}(\eta) -
    \gamma_{n}(\xi)   \gamma_{n+1}(\eta)
  \right].
\end{equation}
Now, using (\ref{d-alpha-symm}) one can rewrite the left-hand 
side of this equation as 
  $(\xi/\eta - 1 )\partial(\eta)\alpha_{n}(\xi)$
which ends the proof of (\ref{prop-1}).

Formula (\ref{prop-1}) is crucial for the derivation of the 
functional representation of the VH. By taking the $\xi \to \eta$ 
limit and using some simple algebra one can come to the very 
important fact: the quantity $f_{n}$ given by
\begin{equation}
  f_{n} = 
  f_{n}\left( \zeta, \mathrm{t} \right) =
  f_{n}\left( \zeta, t_{1}, t_{2}, ... \right) =
  \frac{ \tau_{n-1} }{ \tau_{n-2} }
  \frac{ \alpha_{n}(\zeta) }{ \gamma_{n}(\zeta) }
\end{equation}
satisfies equation
\begin{equation}
   \frac{ 1 }{ \zeta } \, \partial(\zeta) f_{n} =
   \frac{ \partial }{ \partial\zeta } \, f_{n} 
\end{equation}
which means that
\begin{equation}
  f_{n}\left( \zeta, t_{1}, t_{2}, ... \right) =
  F_{n}\left( 
    t_{1} + \zeta, 
    t_{2} + \frac{\zeta^{2}}{2}, 
    t_{3} + \frac{\zeta^{3}}{3}, ... \right)
\label{fn-zeta}
\end{equation}
or, 
\begin{equation}
  f_{n}\left( \zeta, \mathrm{t} \right) =
  F_{n}\left( \mathrm{t} + [\zeta] \right)
\end{equation}
where square brackets indicate the so-called Miwa's shifts:

\begin{equation}
  F\left( \mathrm{t} + \epsilon [\zeta] \right) =
  F\left( ... , t_{k} + \epsilon\frac{\zeta^{k}}{k}, ... \right).
\end{equation}
Setting $\zeta$ equal to zero in (\ref{fn-zeta}) one can get that 
  $F_{n}\left( \mathrm{t} \right) =
  - \tau_{n+1}( \mathrm{t} ) / \tau_{n}( \mathrm{t} ) $
which gives
\begin{equation}
  \frac{ \alpha_{n}(\zeta,\mathrm{t}) }
       { \gamma_{n}(\zeta,\mathrm{t} ) } =
  - 
  \frac{ \tau_{n-2}(\mathrm{t}) }{ \tau_{n-1}(\mathrm{t} ) } 
  \frac{ \tau_{n+1}(\mathrm{t}+[\zeta]) }
       { \tau_{n}  (\mathrm{t}+[\zeta]) }.
\label{fn-zeta-1}
\end{equation}
Noting that (\ref{Delta}) leads to 

\begin{equation}
  \frac{ 1 + \zeta\alpha_{n-1} }{ \gamma_{n} } = 
  \frac{ u_{n-1}\gamma_{n-1} }{ \alpha_{n-1} }
\end{equation}
equation (\ref{fn-zeta-1}) can be rewritten as
\begin{equation}
  \frac{ 1 + \zeta\alpha_{n-1}(\zeta,\mathrm{t}) }
       { \gamma_{n}(\zeta,\mathrm{t} ) } =
  - 
  \frac{ \tau_{n}  (\mathrm{t}) }
       { \tau_{n-1}(\mathrm{t} ) } 
  \frac{ \tau_{n-1}(\mathrm{t}+[\zeta]) }
       { \tau_{n}  (\mathrm{t}+[\zeta]) }.
\label{fn-zeta-2}
\end{equation}
Substitution of (\ref{fn-zeta-1}) and (\ref{fn-zeta-2}) into 
relation (\ref{zcr-gamma-zeta}) leads to the following bilinear 
equation for the tau-functions:
\begin{equation}
  \zeta
  \tau_{n-2}\left(\mathrm{t}\right)
  \tau_{n+1}\left(\mathrm{t}+[\zeta]\right)
  -
  \tau_{n-1}\left(\mathrm{t}\right)
  \tau_{n}  \left(\mathrm{t}+[\zeta]\right)
  +
  \tau_{n}  \left(\mathrm{t}\right)
  \tau_{n-1}\left(\mathrm{t}+[\zeta]\right)
  = 0.
\label{main}
\end{equation}

This is the central formula of this paper. This functional 
equation contains all differential equations of the VH: expanding 
(\ref{main}) in the Taylor series in $\zeta$ 
\begin{equation}
  f(\mathrm{t}+[\zeta]) =
  f(\mathrm{t}) 
  + \zeta\partial_{1}f(\mathrm{t}) 
  + \frac{\zeta^{2}}{2} 
    \left( \partial_{2} + \partial_{11} \right)f(\mathrm{t}) 
  + ...  
\end{equation}
and gathering terms with different powers of $\zeta$ one can 
obtain all Volterra flows. For example, $\zeta^{1}$ terms give
\begin{equation}
  \tau_{n-2} \tau_{n+1} -
  \tau_{n-1} \partial_{1} \tau_{n} + 
  \tau_{n}   \partial_{1} \tau_{n-1} = 0
\end{equation}
which is nothing but equation (\ref{blve-1}). 

To simplify the following formulae hereafter I will also use 
another designation for the Miwa's shifts:
\begin{equation}
  \left( \ep{\zeta}f \right) (\mathrm{t}) =
  f(\mathrm{t} + [\zeta]) 
\end{equation}
and
\begin{equation}
  \left( \en{\zeta}f \right) (\mathrm{t}) =
  f(\mathrm{t} - [\zeta]). 
\end{equation}
In these terms equation (\ref{main}) can be presented as
\begin{equation}
  \zeta \, \tau_{n-2} \, \left( \ep\zeta \tau_{n+1} \right) - 
  \tau_{n-1} \, \left( \ep\zeta \tau_{n} \right) + 
  \tau_{n} \, \left( \ep\zeta \tau_{n-1} \right)
  = 0  
\label{main-pos}
\end{equation}
or
\begin{equation}
  \zeta \, \left( \en\zeta \tau_{n-2} \right) \, \tau_{n+1} - 
  \left( \en\zeta \tau_{n-1} \right) \, \tau_{n} + 
  \left( \en\zeta \tau_{n} \right) \, \tau_{n-1}
  = 0
\label{main-neg}
\end{equation}
(which is equation (\ref{main}) after the shift 
  $ t_{k} \to t_{k} - \zeta^{k}/k $
).

\section{Fay's identities.}  \label{sec-fay}

After we have established the simplest relations describing the Miwa's 
shifts I am going to derive the superposition formulae, i.e. to 
calculate the result of combined action of several Miwa's shifts.

Writing down (\ref{main-pos}) with $\zeta=\xi$, applying 
$\ep\eta$, multiplying the result by 
$\eta\left(\ep\xi\tau_{n-2}\right)$ and then subtracting the 
similar expression with $\xi$ and $\eta$ being interchanged (i.e. 
performing the antisymmetrization with respect to $\xi$ and 
$\eta$) one can get
\begin{equation}
  X_{n}(\xi,\eta) \, \left( \ep\xi\ep\eta \tau_{n} \right) -
  \widetilde{X}_{n}(\xi,\eta) \, \left( \ep\xi\ep\eta \tau_{n-1} \right) 
  = 0
\label{fay-1}
\end{equation}
where
\begin{eqnarray}
  X_{n}(\xi,\eta) &=& 
  \xi  \, \left( \ep\xi\tau_{n}   \right)\left( \ep\eta \tau_{n-1} \right) - 
  \eta \, \left( \ep\xi\tau_{n-1} \right)\left( \ep\eta \tau_{n}   \right) 
\\
  \widetilde{X}_{n}(\xi,\eta) &=& 
  \xi  \, \left( \ep\xi\tau_{n}   \right)\left( \ep\eta \tau_{n-2} \right) - 
  \eta \, \left( \ep\xi\tau_{n-2} \right)\left( \ep\eta \tau_{n}   \right). 
\end{eqnarray}
After the following calculations,
\begin{eqnarray}
  \tau_{n-1} \, \widetilde{X}_{n}(\xi,\eta) &=& 
  \xi  \, \left( \ep\xi\tau_{n} \right)
  \left[ 
    \tau_{n-2} \left( \ep\eta\tau_{n-1} \right)
    - \eta \tau_{n-3} \left( \ep\eta \tau_{n} \right) 
  \right]
\\ &+&
  \eta  \, \left( \ep\eta\tau_{n} \right)
  \left[ 
    \xi \tau_{n-3} \left( \ep\xi \tau_{n} \right) 
    - \tau_{n-2} \left( \ep\xi\tau_{n-1} \right)
  \right]
\\ &=&
  \tau_{n-2} \, X_{n}(\xi,\eta) 
\end{eqnarray}
(where (\ref{main-pos}) with $n \to n-1$ has been used), equation (\ref{fay-1}) reads
\begin{equation}
  \frac{ X_{n}(\xi,\eta) }
       { \tau_{n-1} \left( \ep\xi\ep\eta \tau_{n} \right) }
  = 
  \frac{ X_{n-1}(\xi,\eta) }
       { \tau_{n-2} \left( \ep\xi\ep\eta \tau_{n-1} \right) }
\end{equation}
which means that 
   $X_{n}(\xi,\eta)/\tau_{n-1}\left( \ep\xi\ep\eta \tau_{n} \right)$
is a constant with respect to $n$. Denoting this constant as 
$a(\xi,\eta)$ we come to the two-shift superposition formula
\begin{equation}
  a(\xi,\eta) \, \tau_{n-1} \, \left( \ep\xi\ep\eta \tau_{n} \right) 
  =
  \xi  \, \left( \ep\xi \tau_{n}    \right) \,
          \left( \ep\eta \tau_{n-1} \right) -
  \eta \, \left( \ep\xi \tau_{n-1}  \right) \,
          \left( \ep\eta \tau_{n}   \right).
\label{bip-1}
\end{equation}
By applying (\ref{main-pos}) one can derive from this relation 
the following ones:
\begin{equation}
  a(\xi,\eta) \, \tau_{n-2} \, \left( \ep\xi\ep\eta \tau_{n} \right) 
  = 
  \xi  \, \left( \ep\xi \tau_{n}    \right) \,
          \left( \ep\eta \tau_{n-2} \right) -
  \eta \, \left( \ep\xi \tau_{n-2}  \right) \,
          \left( \ep\eta \tau_{n}   \right)
\label{bip-2}
\end{equation}
\begin{equation}
  a(\xi,\eta) \, \tau_{n-3} \, \left( \ep\xi\ep\eta \tau_{n} \right) 
  = 
  \left( \ep\xi \tau_{n-1}  \right) \,
  \left( \ep\eta \tau_{n-2} \right) -
  \left( \ep\xi \tau_{n-2}  \right) \,
  \left( \ep\eta \tau_{n-1} \right)
\label{bip-3}
\end{equation}
where $a(\xi,\eta)$ is an antisymmetric function 
\begin{equation}
  a(\xi,\eta) = - a(\eta, \xi) 
\end{equation}
which should be determined from the boundary conditions.

Subtraction of equation (\ref{bip-3}) multiplied by $\eta$ from (\ref{bip-1}) gives
\begin{equation}
  a(\xi,\eta) \left[ 
    \tau_{n-1} \, \left( \ep\xi\ep\eta \tau_{n} \right) -
    \eta\tau_{n-2} \, \left( \ep\xi\ep\eta \tau_{n+1} \right) 
  \right] = 
  ( \xi - \eta) \,
  \left( \ep\xi \tau_{n}  \right) \left( \ep\eta \tau_{n-1} \right)
\end{equation}
which can be rewritten as a superposition formula for positive, $\ep\xi$, and 
negative, $\en\eta$, Miwa's shifts. This formula, together with two other ones which 
can also be derived from (\ref{bip-1})--(\ref{bip-3}), are given by
\begin{eqnarray}
  c(\xi,\eta) \, \tau_{n} \, \left( \ep\xi\en\eta \tau_{n} \right) 
  & = &
  \left( \en\eta \tau_{n}  \right) \,
  \left( \ep\xi  \tau_{n} \right) -
  \xi\eta \,
  \left( \en\eta \tau_{n-2}  \right) \,
  \left( \ep\xi  \tau_{n+2} \right)
\label{bin-1}
\\
  c(\xi,\eta) \, \tau_{n+1} \, \left( \ep\xi\en\eta \tau_{n} \right) 
  & = & 
  \left( \en\eta \tau_{n}  \right) \,
  \left( \ep\xi  \tau_{n+1} \right) -
  \xi \,
  \left( \en\eta \tau_{n-1}  \right) \,
  \left( \ep\xi  \tau_{n+2} \right)
\\
  c(\xi,\eta) \, \tau_{n-1} \, \left( \ep\xi\en\eta \tau_{n} \right) 
  & = & 
  \left( \en\eta \tau_{n-1}  \right) \,
  \left( \ep\xi  \tau_{n} \right) -
  \eta \,
  \left( \en\eta \tau_{n-2}  \right) \,
  \left( \ep\xi  \tau_{n+1} \right)
\label{bin-3}
\end{eqnarray}
where $c(\xi,\eta)$ is some symmetric function defined by
\begin{equation}
  c(\xi,\eta) = c(\eta,\xi) = 
  \frac{ \xi - \eta }{ a(\xi,\eta) }.
\end{equation}

It is easy to note that the right-hand side of the antisymmetric 
Fay's identity (\ref{bip-3}) is a $2 \times 2$ determinant. It 
turns out that it can be extended to provide a superposition 
formula for $N$ Miwa's shifts
\begin{equation}
  \det \biggl| \; \ep{\xi_{j}} \, \tau_{n+1-k} \; \biggr|_{j,k=1,...,N}
   =  
  a(\xi_{1}, ... ,\xi_{N})
  \;
  \tau_{n-N} \; ... \; \tau_{n-2} 
  \left( \ep{\xi_{1}} ... \ep{\xi_{N}} \tau_{n+N-1} \right)
\end{equation}
where $a(\xi_{1}, ... ,\xi_{N})$ is given by
\begin{equation}
  a(\xi_{1}, ... ,\xi_{N}) =
  \prod_{1 \le i < j \le N} a(\xi_{i},\xi_{j}).
\end{equation}
In a similar way, the multi-shift analog of (\ref{bip-1}) can be written as
\begin{equation}
  \det \biggl| \; \xi_{j}^{N-k}\ep{\xi_{j}} \, \tau_{n+1-k} \; \biggr|_{j,k=1,...,N}
   =  
  a(\xi_{1}, ... ,\xi_{N})
  \;
  \tau_{n+1-N} \; ... \; \tau_{n-1} 
  \left( \ep{\xi_{1}} ... \ep{\xi_{N}} \tau_{n} \right).
\end{equation}
I give these formulae here without proof, which can be done, say, 
by induction (by reducing the $N \times N$ determinant to $(N-1) 
\times (N-1)$ one and using two-shift formulae (\ref{bip-3}) or 
(\ref{bip-1})).

\section{Baker-Akhiezer function.}  \label{sec-baf}

The aim of this section is to derive (formal) solution of the 
auxiliary problem $\Psi_{n+1} = U_{n} \Psi_{n}$ (see 
(\ref{zcr-system}) and (\ref{zcr-u})), which for a $2$-vector
  $\left( \psi_{n},\varphi_{n} \right)^{T}$ 
can be written as a system 
\begin{eqnarray}
  \psi_{n+1} & = & \lambda \psi_{n} + u_{n} \varphi_{n}
\\
  \varphi_{n+1} & = &  - \psi_{n}
\end{eqnarray}
or as second-order equations
\begin{eqnarray}
&&
  \psi_{n+1} - \lambda \psi_{n} + u_{n} \psi_{n-1} = 0
\label{sps-psi}
\\
&&
  \varphi_{n+1} - \lambda \varphi_{n} + u_{n-1} \varphi_{n-1} = 0.
\label{sps-phi}
\end{eqnarray}
Our starting point are equations (\ref{main-pos}) and (\ref{main-neg}). 
Dividing the later by $\tau_{n-1}\tau_{n}$ it can be presented as 
\begin{equation}
  \frac{ \en\zeta \tau_{n}   }{ \tau_{n} } -
  \frac{ \en\zeta \tau_{n-1} }{ \tau_{n-1} } +
  \zeta u_{n} \frac{ \en\zeta \tau_{n-2} }{ \tau_{n-2} } 
  = 0
\end{equation}
where, recall, 
  $u_{n} = \tau_{n-2} \tau_{n+1} / \tau_{n-1}\tau_{n}$
In a similar way, after the shift $n \to n+1$ (\ref{main-pos})
takes the form
\begin{equation}
  \zeta \frac{ \ep\zeta \tau_{n+2} }{ \tau_{n} } -
  \frac{ \ep\zeta \tau_{n+1} }{ \tau_{n-1} } +
  u_{n} \frac{ \ep\zeta \tau_{n} }{ \tau_{n-2} } = 0.
\end{equation}
Comparing these equations with (\ref{sps-psi}) and (\ref{sps-phi}) 
one can conclude that 
  $\lambda^{n} \en\zeta\tau_{n-1} / \tau_{n-1}$
solves (\ref{sps-psi}) while 
  $\lambda^{-n} \ep\zeta\tau_{n} / \tau_{n-2}$
is a solution of (\ref{sps-phi}) This means that, if $\tau_{n}$ 
is a satisfies (\ref{main-pos}) and (\ref{main-neg}), then the 
matrix $\Psi$,
\begin{equation}
  \Psi_{n} = 
  \pmatrix{ 
  \myfrac{ \en\zeta \tau_{n-1} }{ \tau_{n-1} } \;\lambda^{n} 
  &
  - \myfrac{1}{\lambda}
    \myfrac{ \ep\zeta \tau_{n+1} }{ \tau_{n-1} } \;\lambda^{-n} 
  \cr
  \cr
  - \myfrac{1}{\lambda}
    \myfrac{ \en\zeta \tau_{n-2} }{ \tau_{n-2} } \;\lambda^{n} 
  &
  \myfrac{ \ep\zeta \tau_{n} }{ \tau_{n-2} } \;\lambda^{-n} 
  }
  \qquad
  \qquad
  \zeta = \lambda^{-2}
\end{equation}
solves the discrete auxiliary problem. The determinant 
of the matrix $\Psi$
\begin{equation}
  \det\Psi_{n} = 
  \frac{
    \left( \en\zeta \tau_{n-1} \right)
    \left( \ep\zeta \tau_{n}   \right)
    - \zeta
    \left( \en\zeta \tau_{n-2} \right)
    \left( \ep\zeta \tau_{n+1} \right) }
  { \tau_{n-2}\tau_{n-1} }
\end{equation}
is given by
\begin{equation}
  \det\Psi_{n} = c(\zeta,\zeta) \frac{\tau_{n}}{\tau_{n-2}}
\end{equation}
as follows from (\ref{bin-3}). So, $\det\Psi_{n}$ is non-zero, 
except for some special values of $\zeta$, which means that the 
columns of $\Psi_{n}$ are two linearly independent solutions of 
our scattering problem, i.e. its basis.

Thus we have come to the point which demonstrates one of the main 
advantages of functional representation of the integrable 
hierarchy: using the Miwa's shifts one can express solutions of 
the auxiliary problem (which are the central objects of the 
inverse scattering technique) in terms of the solutions of the 
original equation. In the following sections I use this fact to 
enhance results given by the IST, such as, e.g., the generating 
function for the conservation laws, presenting them in a compact 
form and directly in terms of the tau-functions, not invoking some 
intermediate quantities such as Jost functions or scattering data.

\section{Conservation laws.}  \label{sec-cl}

The Volterra model is an integrable system possessing an infinite 
number of constants of motion. A few first of them can be written 
as
\begin{eqnarray}
  I_{1} & = & \sum u_{n},
  \\
  I_{2} & = & \sum 2 u_{n+1}u_{n} + u_{n}^{2},
  \\
  I_{3} & = &  \sum 
    3 u_{n+1}u_{n}u_{n-1} + 
    3 u_{n+1}u_{n}^{2} + 
    3 u_{n}^{2}u_{n-1} + 
    u_{n}^{3}.
\end{eqnarray}
Let us introduce the conserved densities $U_{n}^{(j)}$ by 
\begin{equation}
  I_{j} = \sum_{n} U_{n}^{(j-1)}.
\label{cl-def-U}
\end{equation}
The first two ones can be presented as
\begin{eqnarray}
  U_{n}^{(0)} & = & u_{n},
  \\
  U_{n}^{(1)} & = & u_{n+1}u_{n} + u_{n}u_{n-1} + u_{n}^{2}.
\end{eqnarray}
In terms of the tau-functions $U_{n}^{(0)}$ is given by
\begin{equation}
  U_{n}^{(0)}  = 
  \frac{ \tau_{n-2} \tau_{n+1} }{ \tau_{n-1} \tau_{n} }
\end{equation}
while the second one, by means of the Volterra equation 
(\ref{blve-1}),
\begin{equation}
  u_{n} = \partial_{1} \ln\frac{ \tau_{n} }{ \tau_{n-1} },
\end{equation}
can be presented as
\begin{equation}
  U_{n}^{(1)}  = 
  \frac{ 1 }{ \tau_{n-1} \tau_{n} }
  \left[
    \tau_{n-2} \, \partial_{1} \tau_{n+1} -
    \tau_{n+1} \, \partial_{1} \tau_{n-2} 
  \right].
\end{equation}
One can note that $U_{n}^{(0)}$ and $U_{n}^{(1)}$ coincide with 
the cofactors of $\zeta^{0}$ and $\zeta^{1}$ in the Taylor 
expansion of the quantity
\begin{equation}
  \left( \en\zeta \tau_{n-2} \right)
  \left( \ep\zeta \tau_{n+1} \right) 
  = 
  \tau_{n-2} \tau_{n+1} +
  \zeta
  \left(
    \tau_{n-2} \, \partial_{1} \tau_{n+1} -
    \tau_{n+1} \, \partial_{1} \tau_{n-2} 
  \right) +
  O\left(\zeta^{2}\right).
\end{equation}
By straightforward (but rather cumbersome) calculations one can 
find that the same occurs for $U_{n}^{(2)}$, $U_{n}^{(3)}$ etc. It 
turns out that this is indeed the case: it can be shown that the 
series
\begin{equation}
  U_{n}(\zeta) = \sum_{j=0}^{\infty} \zeta^{j} \, U_{n}^{(j)} 
\end{equation}
is nothing but 
  $\left( \en\zeta \tau_{n-2} \right)\left( \ep\zeta \tau_{n+1} \right) /
   \tau_{n-1}\tau_{n}$:
\begin{equation}
  U_{n}\left(\zeta, \mathrm{t}\right) =
  \frac{
    \tau_{n-2}\left(\mathrm{t}-[\zeta]\right)
    \tau_{n+1}\left(\mathrm{t}+[\zeta]\right) }
  { \tau_{n-1}\left(\mathrm{t}\right)
    \tau_{n}  \left(\mathrm{t}\right) }.
\end{equation}

The proof of this statement is based on the commutativity of 
differentiating and the Miwa's shifts and exploits the 
superposition formulae (\ref{bin-1})--(\ref{bin-3}) in the 
particular case of $\xi = \eta$:
\begin{eqnarray}
  c(\zeta) \, \tau_{n}^{2}
  & = &
  \left( \en\zeta \tau_{n} \right) \,
  \left( \ep\zeta \tau_{n} \right) -
  \zeta^{2} \,
  \left( \en\zeta \tau_{n-2} \right) \,
  \left( \ep\zeta \tau_{n+2} \right)
\label{cl-1}
\\
  c(\zeta) \, \tau_{n-1} \tau_{n}
  & = & 
  \left( \en\zeta \tau_{n-1} \right) \,
  \left( \ep\zeta \tau_{n}   \right) -
  \zeta \,
  \left( \en\zeta \tau_{n-2}  \right) \,
  \left( \ep\zeta \tau_{n+1} \right)
\label{cl-2}
\end{eqnarray}
where
\begin{equation}
  c(\zeta) = c(\zeta,\zeta).
\end{equation}
To make the following formulae more readable I will use in this 
section the $\pm$ designation for $\ep\zeta$ and $\en\zeta$:
\begin{equation}
  \tau_{n}^{+} = \ep\zeta \tau_{n},
\qquad
  \tau_{n}^{-} = \en\zeta \tau_{n}.
\end{equation}
Expressing from (\ref{blve-1}) $\partial_{1}\tau_{n-2}$ and 
$\partial_{1}\tau_{n+1}$ in terms of $\partial_{1}\tau_{n-1}$ and 
$\partial_{1}\tau_{n}$
\begin{eqnarray}
  \partial_{1}\tau_{n-2} & = & 
  \frac{1}{\tau_{n-1}} 
    \left(\tau_{n-2}\,\partial_{1}\tau_{n-1} - \tau_{n-3}\tau_{n} \right)
  \\
  \partial_{1}\tau_{n+1} & = & 
  \frac{1}{\tau_{n}} 
    \left(\tau_{n+1}\,\partial_{1}\tau_{n} - \tau_{n-1}\tau_{n+2} \right)
\end{eqnarray}
one can obtain for the derivative of 
  $\tau_{n-2}^{-}\tau_{n+1}^{+}$
the following expression:
\begin{equation}
  \partial_{1} \tau_{n-2}^{-}\tau_{n+1}^{+} = 
  \frac{\tau_{n-2}^{-}\tau_{n+1}^{+}}{\tau_{n-1}^{-}\tau_{n}^{+}}
    \; \partial_{1} \tau_{n-1}^{-}\tau_{n}^{+} 
  + \frac{\tau_{n-2}^{-}\tau_{n-1}^{+}\tau_{n+2}^{+}}{\tau_{n}^{+}}
  - \frac{\tau_{n-3}^{-}\tau_{n}^{-}\tau_{n+1}^{+}}{\tau_{n-1}^{-}}
\end{equation}
or,
\begin{equation}
  \tau_{n-1}^{-}\tau_{n}^{+}   \; \partial_{1} \tau_{n-2}^{-}\tau_{n+1}^{+} -
  \tau_{n-2}^{-}\tau_{n+1}^{+} \; \partial_{1} \tau_{n-1}^{-}\tau_{n}^{+} =
  \tau_{n-2}^{-}\tau_{n-1}^{-}\tau_{n-1}^{+}\tau_{n+2}^{+} -
  \tau_{n-3}^{-}\tau_{n}^{-}\tau_{n}^{+}\tau_{n+1}^{+}.
\label{cl-3}
\end{equation}
Using (\ref{cl-2}) the left-hand side of (\ref{cl-3}) can be 
rewritten as 
\begin{eqnarray}
  \mbox{lhs(\ref{cl-3})} & = & 
  c(\zeta)\left(
    \tau_{n-1}\tau_{n}   \; \partial_{1} \tau_{n-2}^{-}\tau_{n+1}^{+} -
    \tau_{n-2}^{-}\tau_{n+1}^{+} \; \partial_{1} \tau_{n-1}\tau_{n} 
  \right)
\nonumber\\[2mm] & = &
  c(\zeta) \; \tau_{n-1}^{2}\tau_{n}^{2} \; 
  \partial_{1} 
  \frac{\tau_{n-2}^{-}\tau_{n+1}^{+}}{\tau_{n-1}\tau_{n}}.
\label{cl-4}
\end{eqnarray}
At the same time, the right-hand side of (\ref{cl-3}), by virtue 
of (\ref{cl-1}), is
\begin{eqnarray}
  \mbox{rhs(\ref{cl-3})} & = & 
  c(\zeta)\left(
    \tau_{n-1}^{2} \tau_{n-2}^{-} \tau_{n+2}^{+} -
    \tau_{n}^{2}   \tau_{n-3}^{-} \tau_{n+1}^{+} 
  \right)
\nonumber\\[2mm] & = &
  c(\zeta) \; \tau_{n-1}^{2}\tau_{n}^{2} \; 
  \left(
    \frac{\tau_{n-2}^{-}\tau_{n+2}^{+}}{\tau_{n}^{2}} -
    \frac{\tau_{n-3}^{-}\tau_{n+1}^{+}}{\tau_{n-1}^{2}}
  \right).
\label{cl-5}
\end{eqnarray}
Comparing (\ref{cl-4}) and (\ref{cl-5}) we come to the following 
result:
\begin{equation}
  \partial_{1} U_{n}\left(\zeta, \mathrm{t}\right) =
  W_{n}\left(\zeta, \mathrm{t}\right) - 
  W_{n-1}\left(\zeta, \mathrm{t}\right)
\end{equation}
where
\begin{equation}
  W_{n}\left(\zeta, \mathrm{t}\right) =
  \frac{
    \tau_{n-2}\left(\mathrm{t}-[\zeta]\right) \,
    \tau_{n+2}\left(\mathrm{t}+[\zeta]\right) }
  { \tau_{n}^{2}\left(\mathrm{t}\right) }.
\end{equation}
This means that $U_{n}\left(\zeta, \mathrm{t}\right)$ is indeed 
the conserved density of the first Volterra flow $\partial_{1}$ 
(i.e. of the Volterra equation (\ref{ve-1})). A little bit more 
lengthy calculations (omitted here) show that the same is valid 
for all Volterra flows:
\begin{equation}
  \sum_{j=1}^{\infty} \eta^{j-1} \partial_{j} \;
  U_{n}\left(\xi, \mathrm{t}\right) =
  W_{n}\left(\xi,\eta, \mathrm{t}\right) - 
  W_{n-1}\left(\xi,\eta, \mathrm{t}\right)
\label{cl-6}
\end{equation}
where
\begin{equation}
  W_{n}\left(\xi,\eta, \mathrm{t}\right) =
  \frac{ 1 }{ c(\eta)c^{2}(\xi,\eta) }
  \frac{
    \tau_{n-2}\left(\mathrm{t}-[\xi]-[\eta]\right) \,
    \tau_{n+2}\left(\mathrm{t}+[\xi]+[\eta]\right) }
  { \tau_{n}^{2}\left(\mathrm{t}\right) }.
\end{equation}

Expanding (\ref{cl-6}) in power series in $\xi$ and $\eta$ one can 
get an infinite number of divergence-like conservation laws 
\emph{for all} equations of the VH:

\begin{equation}
  \partial_{k} \; U_{n}^{(j)} =
  W_{n}^{(j,k)} - W_{n-1}^{(j,k)}.
\label{cl-7}
\end{equation}
Note that the conserved densities $U_{n}^{(j)}$ are the same for 
all Volterra equations (as was expected) while the right-hand side 
of (\ref{cl-7}) depends on which Volterra flow we are dealing 
with.

After summing over all $n$, in the case of proper boundary 
conditions, one gets an infinite number of constants of motion 
$I^{(j)}$ given by (\ref{cl-def-U}) common for all equations of 
the hierarchy,
\begin{equation}
  \partial_{k} \; I_{j} = 0 
\end{equation}
and their generating function

\begin{equation}
  I(\zeta) = \sum_{j=1}^{\infty} \zeta^{j} \, I_{j} 
\end{equation}
which is given by
\begin{equation}
  I(\zeta) = \zeta \sum_{n} U_{n}(\zeta) 
\end{equation}
or, finally,
\begin{equation}
  I(\zeta) =
  \zeta \sum_{n} 
  \frac{
    \tau_{n-2}\left(\mathrm{t}-[\zeta]\right) \,
    \tau_{n+1}\left(\mathrm{t}+[\zeta]\right) }
  { \tau_{n-1}(\mathrm{t}) \tau_{n}(\mathrm{t}) }.
\end{equation}

\section{Backlund transformations.}  \label{sec-bt}

The aim of this section is to discuss the Miura and Backlund 
transforms and to expose their inner structure which becomes 
transparent when we rewrite them in the terms of the Miwa's 
shifts.

Here I will follow the paper \cite{KW} by Kajinaga and Wadati. 
The discrete Miura transformation \cite{W,Volkov} is defined by
\begin{equation}
  u_{n} = \mu^{2} \left( 1 + q_{n-1} \right) \left( 1 - q_{n} \right)
\label{bt-1}
\end{equation}
where $\mu$ is an arbitrary constant parameter. It links solutions 
of the Volterra equation (\ref{ve-1}) $u_{n}$ with solutions of 
the so-called modified Volterra lattice,
\begin{equation}
  \dot q_{n} = 
  \mu^{2} \left( 1 - q_{n}^{2} \right) 
  \left( q_{n+1} - q_{n-1} \right)
\end{equation}
The Miura transformation $u_{n} \to q_{n}$ can be used to 
construct the Backlund transformation $u_{n} \to u'_{n}$ of the 
Volterra equation (and in fact the whole VH). It is easy to verify 
that, if $u_{n}$ solves (\ref{ve-1}) and $q_{n}$ is defined by 
(\ref{bt-1}), then the function
\begin{equation}
  u'_{n} = \mu^{2} \left( 1 - q_{n-1} \right) \left( 1 + q_{n} \right)
\label{bt-2}
\end{equation}
is also a solution of the Volterra equation (and, again, of all 
equations of the VH).

Now let us look at the above construction from the viewpoint of 
the functional representation of the VH. 
By straightforward calculations one can 
show that in terms of the function
\begin{equation}
  q'_{n} =
  1 - 
  2 \frac{ \tau_{n+1} \left(\ep\zeta \tau_{n}   \right) }
         { \tau_{n}   \left(\ep\zeta \tau_{n+1} \right) }
\label{bt-q1}
\end{equation}
equation (\ref{main-pos}) becomes
\begin{equation}
  4\zeta u_{n} = \left( 1 + q'_{n-1} \right) \left( 1 - q'_{n} \right)
\label{bt-4}
\end{equation}
which is exactly (\ref{bt-1}) with $\mu^{2} = 1 / 4\zeta$, i.e 
the right-hand side of (\ref{bt-q1}) is explicit realization of 
the Miura transform. Now the obvious question is to calculate 
$u'_{n}$ given by (\ref{bt-2}). It is easy to see, by 
substituting (\ref{bt-q1}) into the right-hand side of 
(\ref{bt-2}) and using once more relation (\ref{main-pos}) that
\begin{equation}
  \left( 1 - q'_{n-1} \right) \left( 1 + q'_{n} \right) =
  4\zeta \left( \ep\zeta u_{n+1} \right)
\end{equation}
i.e.
\begin{equation}
  u'_{n} = \ep\zeta u_{n+1}. 
\label{bt-u-p}
\end{equation}
This means that the Backlund transform constructed by the recipe 
(\ref{bt-1})--(\ref{bt-2}) with (\ref{bt-q1}) is just the Miwa 
shift combined with the shift $n \to n+1$.

But $q'_{n}$ is not the only solution of (\ref{bt-4}) or 
(\ref{bt-1}). Indeed, starting from (\ref{main-neg}) one can show 
that the function $q''_{n}$ given by
\begin{equation}
  q''_{n} =
  2 \frac{ \left(\en\zeta \tau_{n}   \right) \tau_{n-1} }
         { \left(\en\zeta \tau_{n-1} \right) \tau_{n}   }
  - 1
\label{bt-q2}
\end{equation}
also satisfies
\begin{equation}
  \left( 1 + q''_{n-1} \right) \left( 1 - q''_{n} \right) = 4\zeta u_{n}
\end{equation}
while
\begin{equation}
  \left( 1 - q''_{n-1} \right) \left( 1 + q''_{n} \right) = 
  4\zeta \left( \en\zeta u_{n-1} \right)
\end{equation}
which gives the $\zeta$-transform in the opposite direction
\begin{equation}
  u_{n} \to u''_{n} = \en\zeta u_{n-1} 
\label{bt-u-n}
\end{equation}

However, these transformations are too simple to be interesting: 
they do not change the structure of solutions contrary to, say, 
Backlund-Darboux ones which add solitons. This is clearly seen if 
one acts by (\ref{bt-u-p}) or (\ref{bt-u-n}) on the simplest 
(constant) solution of the VH, $u_{n} = u_{\infty} = 
\mbox{constant}$:
\begin{equation}
  u_{n}' = u''_{n} = u_{n} = u_{\infty}.
\end{equation}

At the same time, one can get more rich transformations by 
\textit{linear superposition} of (\ref{bt-u-p}) and 
(\ref{bt-u-n}) rewritten in terms of the tau-functions, 
\begin{equation}
  \tau_{n} \to \ep\zeta \tau_{n+1} 
  \qquad \mbox{and} \qquad
  \tau_{n} \to \en\zeta \tau_{n-1}. 
\end{equation}
It is non-trivial moment: our equations are nonlinear and in 
general a linear combination of two solutions is not a solution. 
However in our case $\ep\zeta \tau_{n+1}$ and 
$\en\zeta\tau_{n-1}$ are not completely independent and it can be 
shown that the transformation $\tau_{n}\to\widetilde\tau_{n}$ 
defined by
\begin{equation}
  \widetilde\tau_{n}(\mathrm{t}) =
  \tau_{n+1}\left( \mathrm{t} + [\eta] \right) +
  \eta^{-n} \exp\left[ \chi(\eta,\mathrm{t}) \right]
  \tau_{n-1}\left( \mathrm{t} - [\eta] \right) 
\label{bt}
\end{equation}
which leads to the Miura transformation $u_{n} \to q_{n}$ that 
generalizes (\ref{bt-q1}) and (\ref{bt-q2}),
\begin{equation}
  q_{n} =
  1 - 
  2 \frac{ \tau_{n+1} \widetilde\tau_{n-1} }
         { \tau_{n}   \widetilde\tau_{n} }
  =
  2 \eta \frac{ \tau_{n-1} \widetilde\tau_{n+1} }
              { \tau_{n}   \widetilde\tau_{n} }
  - 1,
\end{equation}
is indeed a Backlund transformation: if $\tau_{n}$ solves 
(\ref{main-pos}), so does $\widetilde\tau_{n}$ 
provided
\begin{equation}
  \chi(\eta,\mathrm{t}+[\zeta]) - \chi(\eta,\mathrm{t}) =
  \Gamma(\eta,\zeta) - \Gamma(\eta,0)
\end{equation}
with
\begin{equation}
  \Gamma(\eta,\zeta) =
  \ln\frac{ c(\eta, \zeta) }{ a(\eta, \zeta) }.
\end{equation}
It is easy to see that $\chi$ is a linear function of times
\begin{equation}
  \chi(\eta,\mathrm{t}) =
  \sum_{j=1}^{\infty} \chi_{j}(\eta) \, t_{j}
\end{equation}
with the coefficients $\chi_{j}(\eta)$ being determined by the 
Taylor series for $\Gamma(\eta,\zeta)$ as a function of $\zeta$:
\begin{equation}
  \chi_{j}(\eta) = \frac{1}{j} \; \Gamma_{j}(\eta),
  \qquad
  i=1,2,...
\end{equation}
where 
\begin{equation}
  \sum_{j=0}^{\infty} \Gamma_{j}(\eta) \, \zeta^{j} = 
  \ln\frac{ c(\eta, \zeta) }{ a(\eta, \zeta) }.
\label{disp-law-bt}
\end{equation}
Contrary to (\ref{bt-u-p}) and (\ref{bt-u-n}), 
transformation (\ref{bt}) gives non-trivial results. 
Say, applying it to the 
tau-function corresponding to the constant solution one gets 
nothing but the one-soliton one. The elementary Backlund 
transformations (\ref{bt}) are commutative and can 
be used to construct more complex ones (by superposition). I do 
not discuss this matter further because the main result of this 
section is (\ref{bt}): it has been shown that by means of the 
Miwa's shifts it is possible to derive the \textit{explicit} form 
of the Backlund transformations which can hardly be done in the 
framework of other approaches.

\section{Dark solitons.}  \label{sec-sls}

Now I want to discuss the soliton solutions of the 
VH. After we have presented all differential equations of the 
hierarchy in the functional form (\ref{main-pos}) this problem 
becomes algebraic one and can be solved, as is shown below, by 
means of the elementary matrix calculus giving as a result common 
soliton solutions of all Volterra equations.

In what follows I will restrict myself to the case of the 
so-called finite-density boundary conditions
\begin{equation}
  \lim_{n \to \infty} u_{n} = u_{\infty}. 
\label{bc-u}
\end{equation}
In terms of the tau-functions (\ref{bc-u}) gives

\begin{equation}
  \tau_{n} 
  \sim
  u_{\infty}^{n^{2}/4} e^{n \varphi} 
  \quad 
  \mbox{as}
  \quad 
  n \to \infty.
\end{equation}
Note that a choice of $\varphi$ does not affect the value of 
$u_{n}$ since the latter is invariant under 
  $\tau_{n} \to \tau_{n} q^{n}$
transforms. I will determine this function from the condition that 
the simplest (vacuum) solution is given by

\begin{equation}
  \tau_{n}^{(vac)}(\mathrm{t}) =
  u_{\infty}^{n^{2}/4} e^{n \varphi(\mathrm{t})}. 
\label{tau-vac}
\end{equation}
After substituting (\ref{tau-vac}) into, for example, 
(\ref{main-pos}) one comes to the equation

\begin{equation}
  \exp\left\{ 
    \varphi\left( \mathrm{t} + [\zeta] \right) -
    \varphi\left( \mathrm{t} \right) 
  \right\} = f(\zeta)
\label{varphi-eq}
\end{equation}
where $f(\zeta)$ is the solution of the equation

\begin{equation}
  \zeta u_{\infty} f^{2} - f + 1 = 0
\end{equation}
which satisfies condition $f(0)=1$, i.e.

\begin{equation}
  f(\zeta) 
  =
  \frac{2}{ 1 + \sqrt{ 1 - 4u_{\infty}\zeta }} \, .
\end{equation}
Equation (\ref{varphi-eq}) can be solved by the ansatz

\begin{equation}
  \varphi\left( \mathrm{t} \right) =
  \sum_{j=1}^{\infty} \varphi_{j} \, t_{j}
\end{equation}
where the coefficients $\varphi_{j}$ should be determined from 
the condition
  $\sum_{j}\varphi_{j}\zeta^{j}/j = \ln f(\zeta)$
which, after applying 
$\zeta d / d \zeta $, can be rewritten as

\begin{equation}
  \sum_{j=1}^{\infty} \varphi_{j} \, \zeta^{j} = 
  \frac{1 - \sqrt{ 1 - 4u_{\infty}\zeta }}{ 2 \sqrt{ 1 - 4u_{\infty}\zeta }}
\, .
\end{equation}
Now one can take into account the boundary conditions (\ref{bc-u}) 
by looking for the solution of (\ref{main-pos}) in the form

\begin{equation}
  \tau_{n} =
  \tau_{n}^{(vac)} \, \omega_{n},
  \qquad
  \lim_{n \to \infty} \omega_{n} = 1.
\end{equation}
In terms of $\omega_{n}$ equation (\ref{main-pos}) can be written 
as

\begin{equation}
  \left[ f(\zeta) - 1 \right] 
    \omega_{n-2} \left(\en\zeta\omega_{n+1} \right)
  - f(\zeta) \, 
    \omega_{n-1} \left(\en\zeta\omega_{n}   \right) 
  + \omega_{n}   \left(\en\zeta\omega_{n-1} \right) 
  = 0
\label{omega-eq}
\end{equation}
and namely this equation will be solved in what follows.

I will not present a detailed textbook-like derivation of 
$N$-soliton solutions of (\ref{omega-eq}) but show that they are 
given by

\begin{equation}
  \omega_{n} = \omega\left( A_{n} \right),
  \qquad
  \omega\left( A \right) = \det \left| \unitmatrix + A \right|
\end{equation}
where $\unitmatrix$ is the $N \times N$ unit matrix and $A_{n}$ is 
the matrix with the elements

\begin{equation}
  A_{n}^{(jk)} =
  \frac{ \ell_{j} \, a_{nk}\left(\mathrm{t}\right) }
       { \xi_{j} - \eta_{k} }
  \qquad
  j,k = 1, ... , N
\label{matrices}
\end{equation}
(matrices of this type often arise in the soliton theory). Here 
$\ell_{j}$, $\xi_{j}$ and $\eta_{k}$ are some constant parameters 
and $a_{nk}\left(\mathrm{t}\right)$ are some functions of all 
times $t_{1}$, $t_{2}$, ..., which will be determined below. I 
will rely on the algebraic properties of the matrices 
(\ref{matrices}) satisfying

\begin{equation}
  L A - A R = | \,\ell\, \rangle \langle a |
\end{equation}
where
  $L = \mbox{diag}\left( \xi_{1}, ... , \xi_{N} \right)$,
  $R = \mbox{diag}\left( \eta_{1}, ... , \eta_{N} \right)$
and the bra-ket notation is used for the $N$-columns, 
  $| \,\ell\, \rangle  = \left( \ell_{1}, ... , \ell_{N} \right)^{T}$
and the $N$-rows,
  $\langle a |  = \left( a_{1}, ... , a_{N} \right)$.
In the appendix one can find some basic facts related to such 
matrices together with the derivation of the identity which we 
need for our purpose and which can be formulated as follows. If 
we have an one-parameter family of matrices $M_{\zeta}$
\begin{equation}
  M_{\zeta} = I_{\zeta}J_{\zeta}^{-1}
\label{M-def}
\end{equation}
with
\begin{equation}
  I_{\zeta} = \unitmatrix - \zeta L 
  \quad
  \mbox{and}
  \quad
  J_{\zeta} = \unitmatrix - \zeta R 
\end{equation}
then the determinants of the matrices deformed by means of 
$M_{\zeta}$,
\begin{eqnarray}
  \Omega_{\alpha} & = &
    \det \left| \unitmatrix + AM_{\alpha} \right|,
\label{Omega-def-1}
  \\
  \Omega_{\alpha\beta} & = &
    \det \left| \unitmatrix + AM_{\alpha}M_{\beta} \right|
\label{Omega-def-2}
\end{eqnarray}
satisfy the identity
\begin{equation}
  \alpha ( \beta - \gamma ) \, 
  \Omega_{\alpha}\Omega_{\beta\gamma} 
  +
  \beta ( \gamma - \alpha ) \, 
  \Omega_{\beta}\Omega_{\alpha\gamma} 
  +
  \gamma ( \alpha - \beta ) \, 
  \Omega_{\gamma}\Omega_{\alpha\beta} 
  =  0
\label{Fay}
\end{equation}
which is an elementary version of the famous Fay's formula for 
the theta functions \cite{Fay}.

Now, to solve (\ref{omega-eq}) one has to state that both the Miwa's 
shifts and the shifts of the index $n$ can be obtained by means 
of the $M$-matrices:
\begin{equation}
  \ep\zeta A_{n} = A_{n} M_{\alpha} 
\label{zeta-dep}
\end{equation}
where $\alpha$ is a function of $\zeta$ and
\begin{eqnarray}
  A_{n-1} & = & A_{n} M_{\beta} 
\\ 
  A_{n-2} & = & A_{n} M_{\gamma} 
\end{eqnarray}
where, of course, $M_{\beta}$ and $M_{\gamma}$ should be related 
by $M_{\gamma}=M_{\beta}^{2}$. Then equation (\ref{omega-eq}) can 
be rewritten as

\begin{equation}
  [ f(\zeta) - 1 ] \,
  \Omega_{\alpha}\Omega_{\beta\gamma} 
  +
  \Omega_{\beta}\Omega_{\alpha\gamma} 
  -
  f(\zeta) \,
  \Omega_{\gamma}\Omega_{\alpha\beta} 
  =  0
\label{omega-eq-2}
\end{equation}
(here I used the definitions (\ref{Omega-def-1}) and 
(\ref{Omega-def-2}) with $A=A_{n+1}$) and to solve it one has 
only to determine $\alpha$, $\beta$ and $\gamma$ by comparing it 
with (\ref{Fay}). After some simple algebra one can conclude that 
the proper choice of the parameters is
\begin{equation}
  \alpha(\zeta) = f(\zeta) - 1, 
  \qquad
  \beta = -1,
  \qquad
  \gamma = \infty
\end{equation}
while the restriction $M_{\infty}=M_{-1}^{2}$ leads to
\begin{equation}
  R = L^{-1}.
\end{equation}
This completes the solution of the problem. The final result can 
be rewritten as
\begin{equation}
  \omega_{n} = \det\left| \;
    \delta_{jk} +  
    \frac{ a_{k}(\mathrm{t})\,\xi_{k}^{-n} }{ 1 - \xi_{j}\xi_{k} }
  \; \right|
\end{equation}
(here the 'left' multipliers $\ell_{j}\xi_{j}$ has been 
incorporated into $a_{k}$ by the gauge transform 
  $A_{n} \to S^{-1}A_{n}S$
which does not change the determinants) with
\begin{equation}
  a_{k}(\mathrm{t}) = 
  a_{k}^{(0)}
  \exp\left[ \phi_{k}(\mathrm{t}) \right].
\end{equation}
The time dependence is given by
\begin{equation}
  \phi_{k}(\mathrm{t}) =  \phi\left( \xi_{k}, \mathrm{t} \right)
\end{equation}
with 
\begin{equation}
  \phi\left( \xi, \mathrm{t} \right) =
  \sum_{j=1}^{\infty} \nu_{j}(\xi) \, t_{j}
\end{equation}
where the coefficients $\nu_{j}$ should be determined from the series
\begin{equation}
  \sum_{j=1}^{\infty} \nu_{j}(\xi) \, \frac{ \zeta^{j} }{ j } = 
  \ln \frac{ 1 - \alpha(\zeta)\,\xi }{ 1 - \alpha(\zeta)\,\xi^{-1} }
\label{disp-law-sls}
\end{equation}
which is a condensed form of \textit{all dispersion laws} of the VH in 
the dark soliton case.

\section{Volterra and other hierarchies}   \label{sec-kp}

As the last application of the functional approach I am going to
discuss an interesting question of interrelations between
different integrable systems. The idea behind examples given below
is that starting from the main equation of this work (\ref{main})
one can derive some other equations which turn out to be closely
related to hierarchies different from the VH.

Equation (\ref{main}) is an functional-difference
equation which relates \emph{four} tau-functions with different index
$n$. However by simple algebra one can derive, as its
consequences, equations involving less number of $\tau_{n}$'s. One
of them can be written as a \emph{two}-point relation

\begin{equation}
  D(\zeta) \; \tau_{n} \cdot \tau_{n-1} =
  \left(\en\zeta \tau_{n-1} \right) \left(\ep\zeta \tau_{n} \right) -
  \tau_{n-1}\tau_{n}
\label{mkph}
\end{equation}
where $D(\zeta)$ is a "hierarchical" version of the Hirota's
bilinear operators:
  $D(\zeta) = \sum_{j=1}^{\infty} \zeta^{j} D_{j}$
(all formulae of this section are written for the case of zero 
boundary conditions).
Expanding this relation in the multidimensional bilinear Taylor 
series using the identity
\begin{equation}
  \left(\ep\zeta a \right) \left(\en\zeta b \right) =
    F\left( \zeta, D \right) \; a \cdot b
\end{equation}
where the bilinear operator $F$ is given by
\begin{equation}
  F\left( \zeta, D \right) =
  1 + \zeta D_{1} +
  \frac{ \zeta^{2} }{ 2 } \left( D_{2} + D_{11} \right) +
  \frac{ \zeta^{3} }{ 6 } \left( 2D_{3} + 3D_{21} + D_{111} \right) +
  ...
\end{equation}
(here $D_{j\,k...} = D_{j}D_{k}...$) one can get an infinite set
of differential equations for the pair $\tau_{n}$ and
$\tau_{n-1}$. A few first of them are
\begin{eqnarray}
&&
  \left( - D_{2} + D_{11} \right) \;
    \tau_{n} \cdot \tau_{n-1} = 0
\\&&
  \left( - 4D_{3} + 3D_{21} + D_{111} \right) \;
    \tau_{n} \cdot \tau_{n-1} = 0
\\&&
  \left( - 18D_{4} + 8D_{31} + 3D_{22} + 6D_{211} + D_{1111}\right) \;
    \tau_{n} \cdot \tau_{n-1} = 0
\\&& \qquad\cdots
\nonumber
\end{eqnarray}
Comparing these equations with the ones presented in \cite{JM} one
can conclude that i) equation (\ref{mkph}) can be viewed as the
Miwa's representation for the 1st modified KP hierarchy (according
to the classification of \cite{JM}) and that ii) this hierarchy
can be 'embedded' into the VH in the sense that any tau-function of 
the VH is at the same time a solution of all modified KP equations.

Continuing the procedure of decreasing the number of
tau-functions involved in our functional relations it is possible
to derive the following one:
\begin{equation}
  \frac{\zeta}{2} \;
  D(\zeta) D_{1} \; \tau_{n} \cdot \tau_{n} =
    \left(\en\zeta\tau_{n}\right)\left(\ep\zeta\tau_{n}\right) -
    \tau_{n}^{2}
\end{equation}
which contains only \emph{one} tau-function and which can be
rewritten as
\begin{equation}
  \left[
    \frac{\zeta}{2} D(\zeta)D_{1} + 1 - F(\zeta,D)
  \right] \;
  \tau_{n} \cdot \tau_{n} = 0
\label{kph}
\end{equation}
where operator F was defined above. The simplest equation of this
infinite set,
\begin{equation}
  \left( - 4D_{31} + 3D_{22} + D_{1111} \right) \;
    \tau_{n} \cdot \tau_{n} = 0,
\end{equation}
is nothing but the KP equation. So we have come to an interesting
fact that the KP equation can be 'embedded' into the VH and have
obtained the functional representation (\ref{kph}) for the KP
hierarchy.

\section{Conclusion.}

In this paper I have derived the so-called functional representation of 
the VH. The VH which is an infinite set of differential equations was 
presented as a difference (or functional) one. In section \ref{sec-sls} 
it was demonstrated how this approach can simplify the problem of finding 
the soliton solutions. Another example, which was not discussed here, is 
the derivation of the quasi-periodical ones. To do this one does not need 
now to develop the algebro-geometric scheme or to solve the inverse 
Jacobi problem. Instead one can use the straightforward method (applied, 
e.g. in \cite{V1999} for the Ablowitz-Ladik and in \cite{V2000} for the 
Hirota's bilinear discrete equations). By comparing equation (\ref{main}) 
with the original Fay's identity for the theta functions \cite{Fay} one 
can immediately establish the structure of the solution and determine 
the dispersion law by expanding in series some functions appearing in the 
latter in a way similar to one used in sections \ref{sec-bt} and 
\ref{sec-sls} (see (\ref{disp-law-bt}) an (\ref{disp-law-sls})).

However the main result of this work, to my opinion, is the 
'eliminating' the intermediate objects of the IST, such as Jost or 
Baker-Akhiezer functions. The fact that we have expressed the Backlund 
transforms and the generating function for the conservation laws in terms 
of the solutions of the Volterra equations themselves is interesting not 
only from the theoretical viewpoint. Say, these results can be useful in 
tackling the non-integrable problems related to the Volterra model by 
improving the averaging methods (exploiting the conservation laws) or the 
perturbation theories (which use the Jost functions to construct the 
inverse of the operator appearing after linearization of our 
nonlinear equations).

\section*{Appendix.}
\renewcommand{\theequation}{A.\arabic{equation}}

The aim of this appendix is to derive the identity (\ref{Fay}) 
which can be rewritten as
\begin{equation}
  \sum_{c.p.} 
  \alpha_{i} \left( \alpha_{j} - \alpha_{k} \right)
  \omega_{\alpha_{i}} \omega_{\alpha_{j}\alpha_{k} }
  = 0
\label{Fay-ijk}
\end{equation}
where $\sum_{c.p.}$ is the sum over the cyclic permutations of 
the indices $i,j,k$.

To do this I am going first to derive some formulae relating 
determinants $\omega(B)$ and $\omega(C)$ with 
$\omega(BM_{\zeta}^{-1})$ and $\omega(CM_{\zeta})$ where 
$M_{\zeta}$ is given by (\ref{M-def}) and $B$, $C$ are matrices 
of the type (\ref{matrices}):
\begin{equation}
  L B - B R = | \,\ell\, \rangle \langle b \, |,
\qquad
\qquad
  L C - C R = | \,\ell\, \rangle \langle c \, |.
\label{matrices-BC}
\end{equation}
This can be done as follows. For any matrix (\ref{matrices-BC}) 
and any number $\zeta$ it is easy to derive the chain of 
identities 
\begin{equation}
  BJ_{\zeta} (\unitmatrix + B) =
  BJ_{\zeta} - I_{\zeta}B + (I_{\zeta} + BJ_{\zeta})B =
  BJ_{\zeta} - I_{\zeta}B + 
  \left(\unitmatrix + BM_{\zeta}^{-1} \right) I_{\zeta}B
\end{equation}
which leads to
\begin{equation}
  \left(\unitmatrix + BM_{\zeta}^{-1} \right) I_{\zeta}BF(B) =
  BJ_{\zeta} + (I_{\zeta}B - BJ_{\zeta}) F(B)
\end{equation}
with
\begin{equation}
  F\left( B \right) = \left( \unitmatrix + B \right)^{-1}
\end{equation}
and
\begin{eqnarray}
  \left( \unitmatrix + BM_{\zeta}^{-1} \right)
  \left( \unitmatrix - I_{\zeta}BF(B)Y \right) & = & 
  \unitmatrix + BM_{\zeta}^{-1} - BJ_{\zeta}Y 
  + (BJ_{\zeta} - I_{\zeta}B)F(B)Y
\\ & = &
  \unitmatrix + BJ_{\zeta} \left(I_{\zeta}^{-1} - Y \right) 
  + (BJ_{\zeta} - I_{\zeta}B)F(B)Y
\label{app-eq-16}
\end{eqnarray}
where $Y$ is an arbitrary matrix. Choosing
\begin{equation}
  Y = Y_{\zeta} = I_{\zeta}^{-1} - J_{\zeta}^{-1}D
\end{equation}
where $D$ is a diagonal matrix, 
  $D = \mbox{diag} \left( c_{j} / b_{j} \right)$
(for simplicity I restrict myself to the case $b_{j} \ne 0$), one 
can continue (\ref{app-eq-16}) as
\begin{eqnarray}
  \left( \unitmatrix + BM_{\zeta}^{-1} \right)
  \left( \unitmatrix - I_{\zeta}BF(B)Y_{\zeta} \right) & = & 
  \unitmatrix + C + (BJ_{\zeta} - I_{\zeta}B)F(B)Y_{\zeta}
\\ & = &
  \left[ 1 + (BJ_{\zeta} - I_{\zeta}B)F(B)Y_{\zeta}F(C) \right] 
  ( \unitmatrix + C ).
\end{eqnarray}
Taking the determinant one can get 
\begin{equation}
  \omega\left(BM_{\zeta}^{-1} \right)
  \det\left| \unitmatrix - I_{\zeta}BF(B)Y_{\zeta} \right| = 
  \omega(C)
  \det\left| \unitmatrix + (BJ_{\zeta} - I_{\zeta}B)F(B)Y_{\zeta}F(C) \right|.
\label{app-det-1}
\end{equation}
From the other hand,
\begin{eqnarray}
  \det\left| \unitmatrix  - I_{\zeta}BF(B)Y_{\zeta} \right| & = & 
  \frac{1}{\omega(B)}
  \det\left| \unitmatrix + B - Y_{\zeta}I_{\zeta}B \right| 
\\ & = & 
  \frac{1}{\omega(B)}
  \det\left| \unitmatrix + B(1 - Y_{\zeta}I_{\zeta}) \right| 
\\ & = & 
  \frac{1}{\omega(B)}
  \det\left| \unitmatrix + CM_{\zeta} \right| 
\\ & = & 
  \frac{\omega(CM_{\zeta})}{\omega(B)}
\label{app-det-2}
\end{eqnarray}
As to the determinant appearing in the right-hand side of 
(\ref{app-det-1}), it can be calculated using the identity
\begin{equation}
  B J_{\zeta} - I_{\zeta} B = \zeta | \,\ell\, \rangle \langle b \, |
\end{equation}
which follows from (\ref{matrices-BC}) and the identity
\begin{equation}
  \det \left| 
    \unitmatrix  + 
    | \, u \, \rangle \langle v | \;
  \right| =
  1 + \langle v | \, u \, \rangle. 
\end{equation}
This leads to 
\begin{equation}
  \det\left| 
    \unitmatrix  + 
    \left( BJ_{\zeta} - I_{\zeta}B \right) 
    F(B) Y_{\zeta}F(C) 
  \right| 
  = 
  1 + 
  \zeta\langle b  | F(B) Y_{\zeta}F(C) |\ell \rangle. 
\label{app-det-3}
\end{equation}
Gathering (\ref{app-det-2}) and (\ref{app-det-3}) we come to
\begin{equation}
  \frac{ \omega\left( BM_{\zeta}^{-1} \right)\omega(CM_{\zeta}) }
       { \omega(B)\omega(C) } 
  = 1 + \zeta \langle b | \, F(B) Y_{\zeta} F(C) | \,\ell\, \rangle .
\end{equation}

Taking this formula with 
\begin{equation}
  B = A M_{\alpha}M_{\beta}M_{\gamma},
  \qquad
  C = A
  \qquad
  \mbox{and}
  \qquad
  \zeta = \alpha
\end{equation}
one can present 
  $\Omega_{\alpha}\Omega_{\beta\gamma} / \Omega\Omega_{\alpha\beta\gamma}$
as
\begin{eqnarray}
  \frac{ \Omega_{\alpha}\Omega_{\beta\gamma} }
       { \Omega\Omega_{\alpha\beta\gamma} }
  & = & 
  \frac{ 
    \omega\left( BM_{\alpha}^{-1} \right)
    \omega\left( CM_{\alpha} \right) } 
    { \omega(B)\omega(C) }
\\ & = & 
  1 + 
  \alpha\langle b | \, F(B) 
  \left[ 
    I_{\alpha}^{-1} -
    J_{\alpha}^{-1} M_{\alpha}^{-1}M_{\beta}^{-1}M_{\gamma}^{-1}
  \right]
  F(A) | \,\ell\, \rangle 
\\ & = & 
  1 + 
  \alpha\langle b | \, F(B) 
  I_{\alpha}^{-1}I_{\beta}^{-1}I_{\gamma}^{-1}
  \left[ I_{\beta}I_{\gamma} - J_{\beta}J_{\gamma} \right]
  F(A) | \,\ell\, \rangle. 
\end{eqnarray}
Here, of course, the row $\langle b |$ and the matrix $B$ depend 
on $\alpha$, $\beta$ and $\gamma$, but depend 
\textit{symmetrically} (so this dependence will not be indicated 
explicitly). Using the definition of $I_{\zeta}$ and $J_{\zeta}$ 
one can get
\begin{equation}
  I_{\beta}I_{\gamma} - J_{\beta}J_{\gamma} =
  (\beta + \gamma)(R - L) +
  \beta\gamma \left( L^{2} - R^{2} \right)
\end{equation}
which gives
\begin{equation}
  \Omega_{\alpha}\Omega_{\beta\gamma} =
  u + \alpha(\beta + \gamma)v
\end{equation}
where, again, $u$ and $v$ depend on $\alpha$, $\beta$ and 
$\gamma$, 
\begin{eqnarray}
  u & = & 
  \Omega\,\Omega_{\alpha\beta\gamma} 
  \left\{  
    1 + 
    \alpha\beta\gamma \langle b | \, 
      F(B) I_{\alpha}^{-1}I_{\beta}^{-1}I_{\gamma}^{-1}
      \left( L^{2} - R^{2} \right)
      F(A) 
    | \,\ell\, \rangle 
  \right\},
\\
  v & = & 
  \Omega\,\Omega_{\alpha\beta\gamma} \,
  \langle b | \, 
      F(B) I_{\alpha}^{-1}I_{\beta}^{-1}I_{\gamma}^{-1}
      \left( R - L \right)
      F(A) 
  | \,\ell\, \rangle 
\end{eqnarray}
but depend \textit{symmetrically}. Now it is easy to finish the 
derivation of (\ref{Fay-ijk}): 
\begin{eqnarray}
  \sum_{c.p.} 
  \alpha_{i} \left( \alpha_{j} - \alpha_{k} \right)
  \Omega_{\alpha_{i}} \Omega_{\alpha_{j}\alpha_{k} }
  & = & 
  u \sum_{c.p.} 
    \alpha_{i} \left( \alpha_{j} - \alpha_{k} \right) +
  v \sum_{c.p.} 
    \alpha_{i}^{2} \left( \alpha_{j}^{2} - \alpha_{k}^{2} \right)
\\  & = & 0
\end{eqnarray}

Of course, the calculations presented here hardly have 
anything specific for the theory of integrable systems: 
this is just an exercise in elementary matrix calculus. 
However the bilinear matrix identities of this kind are 
very useful in analysis of integrable equations, giving, 
for example, a possibility to derive the soliton 
solutions in a very easy and short way.


\end{document}